\documentclass[preprint,aps ,nofootinbib]{revtex4}
\usepackage{graphicx}%Include figure files
\usepackage{amsmath}
\usepackage{amsfonts}
\usepackage{amssymb}
\usepackage{color}%
\usepackage{dcolumn}% Align table columns on decimal point
\providecommand{\U}[1]{\protect\rule{.1in}{.1in}}

\newcommand{\f}{\begin{equation}}
\newcommand{\ff}{\end{equation}}
\newcommand{\fa}{\begin{eqnarray}}
\newcommand{\ffa}{\end{eqnarray}}

\begin{document}
\title{ Dynamics of tachyon field in spatially curved FRW universe}
\author{Jian-Li Li}
\email{lijianli1985@163.com}
\author{Jian-Pin Wu}
\email{jianpinwu@yahoo.com.cn} \affiliation{ Center for Relativistic
Astrophysics and High Energy Physics, Department of Physics,
Nanchang University, 330031, China}

\begin{abstract}
The dynamics of a tachyon field plus a barotropic fluid is
investigated in spatially curved FRW universe. We perform a
phase-plane analysis and obtain scaling solutions accompanying
with a discussion on their stability. Furthermore, we construct
the form of scalar potential which may give rise to stable
solutions for spatially open and closed universe separately.

\end{abstract}
\maketitle

\section{Introduction}

The nature and the origin of dark energy is a fundamental puzzle
in modern cosmology. Most dark energy models can be constructed by
using a slowly rolling canonical scalar field, termed as
quintessence. However, there has been increasing interests in
alternative models with a non-canonical kinetic term. Among these
models the most general formalism perhaps is k-essence
\cite{k-essence}. A more specific case is the tachyon field
\cite{tachyon}, which is motivated from string theory. It can be
viewed as a special case of k-essence with Dirac-Born-Infeld (DBI)
action \cite{DBI}. Although the tachyon is an unstable field, its
state parameter in the equation of state varies smoothly between
$-1$ and $0$, thus many authors have already considered the
tachyon field as a suitable candidate for a viable model of dark
energy phenomenologically
\cite{tachyondeTP,tachyondeFP,tachyondeEJC,tachyondeBGTNJW,tachyondeST,tachyondeYG,tachyondeJM,tachyondeAAS,tachyondeBGJW,tachyondeGC}.
For a review, we can refer to Ref.\cite{DERE}.

However, the dynamical dark energy models driven by a scalar field
suffer from the so-called fine-tuning problem and coincidence
problem. In order to address these problems, one may employ scalar
field models exhibiting scaling solutions
\cite{SS1,SS2,SS31,SS32,SS33,SS34,SS35,SS36,SS37,SS38,SS39,SS310,SS311,SS312,SS313,SS314,SS315,tachyondeYG,SS4,tachyondeST,SS6,SS7}.
The scaling solutions as dynamical attractors can considerably
alleviate these two problems. Furthermore, by investigating the
nature of scaling solutions, one can determine whether such
behavior is stable or just a transient feature and explore the
 asymptotic behavior of the scalar field potential.

Many authors have investigated a lot of scalar field models
containing scaling solutions. For instance, a canonical scalar
field with an exponential potential has scaling attractor
solutions \cite{SS2}. For quintessence dark energy model, there
are two scaling solutions. One is fluid-scalar field scaling
solution, which remains subdominant for most of the cosmic
evolution. It is necessary that the scalar field mimics the
background energy density (radiation/matter) in order to respect
the nucleosynthesis constraint and can also alleviate the
fine-tuning problem of initial conditions. The other is scalar
field dominated scaling solution, which is a late time attractor
and gives rise to the accelerated expansion. Since the
fluid-scalar field scaling solution is non-accelerating, we need
an additional mechanism exit from the scaling regime so as to
enter the scalar field dominated scaling solution at late times.
For the discussion on the exiting mechanism, we can refer to Refs.
\cite{DERE,exitingTB,exitingVS,exitingAA,exitingAAC,exitingSAK}.

For tachyon field dark energy, the scaling solutions have also
been investigated by many authors, for example
Refs.\cite{tachyondeTP,tachyondeFP,tachyondeEJC,tachyondeBGTNJW,tachyondeST,tachyondeYG,tachyondeJM,tachyondeAAS,tachyondeBGJW,tachyondeGC}.
To be considered as a realistic model of dark energy, it is found
that the fluid-scalar field scaling solutions are absent and only
the scalar field dominated scaling solutions exist
\cite{tachyondeEJC}. This is very different from the quintessence
case. Therefore, just as pointed out in \cite{tachyondeGC},
tachyon models require more fine-tuning to be consistent with
observations. Nevertheless, it is worthwhile to point out that the
fluid-scalar field scaling solutions may be obtained when the
Gauss-Bonnet coupling between tachyon field and fluid is
considered in the tachyon dark energy model\cite{SS7}.

Although the latest results of WMAP5 have placed a constraint,
$-0.063<\Omega_{k}<0.017$ \cite{WMAP5}, on the flatness of our
observable universe, indicating that our observable universe is
very close to flatness, it is still possible that our observable
universe is spatially curved. Therefore, it is also interesting to
investigate the dynamical behavior of dark energy models in a
spatially curved FRW universe. Recently Copeland {\it et.al.} have
extended such investigations to the quintessence model in
spatially curved FRW universe\cite{CSQ}, following the strategy
they have developed in Ref.\cite{SS312}. Sen and Devi
\cite{tachyondeAAS} have also explored the scaling solutions with
tachyon in modified gravity model employing the same method. In
this Letter, we will closely follow this route to investigate the
dynamics of tachyon dark energy model in spatially curved
universe.

Our Letter is organized as follows. In Section II we present the
associated equations of motion for the tachyon field including the
background fluid and obtain the scaling solutions. Then, we
analyze the stability of these solutions. In Section III we turn
to construct the scalar potential leading to such scaling
solutions. In particular, its asymptotical forms are obtained in
various circumstances for spatially open and closed universe
respectively.

\section{dynamics of the tachyon field in the presence of a fluid}

Let us start with a cosmological model in which the universe is
filled with a tachyon field $\phi$ evolving with a positive
potential $V(\phi)$ and a barotropic fluid with an equation of state
$p_{\gamma}=(\gamma-1)\rho_{\gamma}$, where $\gamma$ is the
adiabatic index. We note that $\gamma=1$ for a pressureless dust and
$\gamma=4/3$ for radiation. The pressure and the energy densities of
the tachyon field $\phi$ are respectively given by
\begin{equation}\label{ab}
  p_{\phi}=-V(\phi)\sqrt{1-\dot{\phi}^{2}} ,
\end{equation}
\begin{equation}\label{ac}
  \rho_{\phi}=\frac{V(\phi)}{\sqrt{1-\dot{\phi}^{2}}},
\end{equation}
then the effective adiabatic index of the tachyon field is given
by

\begin{equation}\label{ad}
\gamma_{\phi}=\frac{\rho_{\phi}+p_{\phi}}{\rho_{\phi}}=\dot{\phi}^{2}.
\end{equation}
From the above equation, we can see that $ 0 < \gamma_{\phi}<1$.

As shown in Ref.\cite{tachyondeAAS}\cite{SS312}\cite{CSQ}, either
in a spatially curved FRW universe or a flat universe but
described by modified gravity models, the Friedmann equation of
the universe can be uniformly written as an effective form
\begin{equation}\label{aa}
  H^{2}=\frac{\rho}{3}L^{2}(\rho(a)),
\end{equation}
where we have set $8\pi/m_{p}=1$. $H\equiv\dot{a}/a$ is the Hubble
parameter and $\rho(a)$ is the total energy density of the universe. In
this paper, we adopt the notion $H=\epsilon\sqrt{H^{2}}$, where
$\varepsilon=+1$ for expanding universe and $\varepsilon=-1$ for
contracting universe. The function $L(\rho(a))$ is assumed to be
positive-definite without loss of generality. We also note that
when $L(\rho(a))=1$, the effective Friedmann equation (\ref{aa})
can be reduced to the standard spatially flat case.

The energy conserved equation of matter and tachyon field are
respectively
\begin{equation}\label{ae}
  \dot{\rho_{\gamma}}+3\gamma H \rho_{\gamma}=0,
\end{equation}
\begin{equation}\label{af}
  \frac{\ddot{\phi}}{1-\dot{\phi}^{2}}+3H\dot{\phi}+\frac{V_{,\phi}}{V}=0.
\end{equation}

Now we intend to demonstrate a phase-plane analysis on this system.
As usual, we define
\begin{equation}\label{ah}
X=\dot{\phi}, \ \ \ \ \ \
Y\equiv\frac{\sqrt{V(\phi)}}{\sqrt{\rho}},
\end{equation}
\begin{equation}\label{al}
\lambda\equiv-\frac{V_{,\phi}}{LV^{3/2}},\ \ \ \ \ \ \Gamma\equiv
V\frac{V_{,\phi\phi}}{V_{,\phi}^{2}},
\end{equation}
where $\rho$ is the total energy density of the universe and the
subscript $N$ denotes the derivative with respect to the number of
e-folds, $N\equiv\ln a$. As a result,
Eq.(\ref{aa}),(\ref{ae})and(\ref{af}) can be rewritten in the form
\begin{equation}\label{ai}
X_{,N}=(X^{2}-1)(3X-\epsilon\sqrt{3}Y\lambda),
\end{equation}

\begin{equation}\label{aj}
Y_{,N}=\frac{Y}{2}[-\epsilon\sqrt{3}XY\lambda+3\gamma-\frac{3Y^{2}(\gamma-X^{2})}{\sqrt{1-X^{2}}}],
\end{equation}

\begin{equation}\label{ak}
\lambda_{,N}=-\sqrt{3}\epsilon
XY\lambda^{2}(\Gamma-\frac{3}{2})-3\lambda[\frac{(\gamma-X^{2})Y^{2}}{\sqrt{1-X^{2}}}-\gamma]\rho\frac{\partial\ln
L}{\partial\rho},
\end{equation}
 and the constraint equation for the total energy density becomes
\begin{equation}\label{ag}
  \frac{Y^{2}}{\sqrt{1-X^{2}}}+\frac{\rho_{\gamma}}{\rho}=1.
\end{equation}
Since $0\leq\frac{Y^{2}}{\sqrt{1-{X}^{2}}}\leq1$, the allowed
range of $X$ and $Y$ is $0\leq X^{2}+Y^{4}\leq1$, namely $0\leq
Y^{2}\leq1$ and $0\leq X^{2}\leq1$.

We consider the special case $\lambda=const$. Taking a derivative
with respect to the scalar field $\phi$, then from Eq.(\ref{al}) we
 obtain
\begin{equation}\label{am}
 \Gamma=\frac{3}{2}+\frac{d\ln L}{d\ln V}.
\end{equation}
The fixed points for this system can be obtained by setting
$X_{,N}=0$ and $Y_{,N}=0$ in Eqs. (\ref{ai}) and (\ref{aj}).
Essentially we have four fixed points:
\begin{equation}\label{an}
X_{c} = 0 ,   Y_{c} = 0,
\end{equation}
\begin{equation}\label{ao}
X_{c} = \pm 1 ,   Y_{c} = 0,
\end{equation}
\begin{equation}\label{ap}
X_{c} =\frac{\lambda}{\sqrt{3}}Y_{s} , Y_{c} =Y_{s},
\end{equation}
\begin{equation}\label{aq}
X_{c} = \pm\sqrt{\gamma} ,   Y_{c} =
\pm\frac{\sqrt{3\gamma}}{\lambda},
\end{equation}
where $Y_{s}$ is defined by
\begin{equation}\label{dp}
Y_{s}=\sqrt{\frac{\sqrt{\lambda^{4}+36}-\lambda^{2}}{6}}.
\end{equation}

We now investigate the stability around the critical points by
evaluating the eigenvalues of the matrix ${\cal M}$. The way of
evaluating eigenvalues has been
 given in \cite{DERE}\cite{SS2}\cite{fff}. Here we list our
 analysis corresponding to each solution as follows and then summarize the
 results in Table I.

\begin{enumerate}
\item For the fluid dominated solution$(X_{c}=0,Y_{c}=0)$, the
eigenvalues are $\mu_{1}=-3,\mu_{2}=3\gamma/2$. Therefore, this
critical point is an unstable saddle point for $\gamma>0$, whereas
it is a stable node for $\gamma=0$ in an expanding universe
$(\epsilon=1)$. However, since the fixed point leads to the density
parameter $\Omega_{\phi}\equiv\frac{\rho_{\phi}}{\rho}=0$, it cannot 
be used as a late-time attractor.

\item The fixed points $(X_{c} =\pm 1 $ $,Y_{c} = 0 )$ correspond
to the scalar field kinetic dominated solution with
$\gamma_{\phi}=\dot{\phi}^{2}$. The eigenvalues are $ \mu_{1}=6,$
$\mu_{2}=3\gamma/2$, indicating that this point is unstable in an
expanding universe. However, in a contracting universe
$(\epsilon=-1)$, it implies unconditional stability when these
solutions exist.

\item For the well-known solution
$({X}_{c}=\lambda Y_{s}/\sqrt{3},$ ${Y}_{c}=Y_{s})$, we have
eigenvalues $ \mu_{1}=-3+\frac{\lambda^{2}}{2}Y_{s}^2 ,$ $
\mu_{2}=-3\gamma+\lambda^{2}Y_{s}^{2}$, where satisfies
$-3\leq\mu_{1}<-3/2$. Moreover, we have $\mu_{2}\leq0$ for $\gamma
\geq \gamma_{s} \equiv \frac{\lambda^{2}}{3}Y_{s}^{2}$, indicating
that, in an expanding universe, the fixed point is stable for
$\gamma\geq\gamma_{s}$, whereas it is an unstable saddle point for
$\gamma<\gamma_{s}$. On the contrary, in a contracting universe the
critical point is unstable for $\gamma\geq\gamma_{s}$ and be a
saddle point for $\gamma<\gamma_{s}$. Since in this solution, the
kinetic energy of the tachyon field is proportional to the potential
energy, we call this solution as the scalar field dominated scaling
solution as Ref.\cite{CSQ}.

\item The value $\gamma_{\phi}$ at the last critical points
$({X}_{c} = \pm\sqrt{\gamma} ,$ ${Y}_{c} =
\pm\frac{\sqrt{3\gamma}}{\lambda} )$ is $\gamma_{\phi}=\gamma$,
which means both energy densities $\rho_{\phi}$ and $\rho_{\gamma}$
decrease with the same rate. The eigenvalues are $
\mu_{1,2}=\frac{3}{4}[\gamma-2\pm\sqrt{17\gamma^{2}-20\gamma+4+\frac{48}{\lambda^{2}}\gamma^{2}\sqrt{1-\gamma}}]
$. From Eq.(\ref{ag}) we have $
0\leq\gamma\leq\gamma_{s}=\frac{\lambda^{2}}{3}Y_{s}^{2},$  note
that $\gamma_{s}$ is always smaller than 1. The real parts of
$\mu_{1}$ and $\mu_{2}$ are both negative under the condition
$\gamma\leq\gamma_{s}$. Obviously, it is an unstable solution in a
contracting universe, whereas in an expanding universe the scaling
solution is always stable. However, we need to caution that the
existence of the scaling solution requires the condition
$0\leq\gamma\leq\gamma_{s}<1$, which is not satisfied for known
realistic fluids. So it is always treated as an unpractical
solution.
\end{enumerate}

\begin{widetext}
\begin{table} [h!]
\begin{tabular} { |c|c|c|c|c|c|c|c|c|c|c|c|c| }
\hline
\raisebox{0.1ex} & \multicolumn{2}{|c|}{$X=0$} & \multicolumn{2}{|c|}{$X=1$} & \multicolumn{2}{|c|}{$X=-1$} & \multicolumn{2}{|c|}{$X=\frac{\lambda}{\sqrt{3}}Y_{s}$} & \multicolumn{2}{|c|}{$X=\sqrt{\gamma}$} & \multicolumn{2}{|c|}{$X=-\sqrt{\gamma}$}  \\
 & \multicolumn{2}{|c|}{$Y=0$} & \multicolumn{2}{|c|}{$Y=0$} & \multicolumn{2}{|c|}{$Y=0$} & \multicolumn{2}{|c|}{$Y=Y_{s}$}& \multicolumn{2}{|c|}{$Y=\frac{\sqrt{3\gamma}}{\lambda}$} & \multicolumn{2}{|c|}{$Y=-\frac{\sqrt{3\gamma}}{\lambda}$}  \\
\hline
 & exists & stable& exists & stable & exists & stable & exists & stable & exists & stable & exists & stable \\
\hline $\epsilon=+1$&$\forall\lambda,\forall\gamma$& No & $\forall
\lambda, \, \forall \gamma$ & No &
 $\forall \lambda, \, \forall \gamma$& No& $\forall\lambda,\forall\gamma$ & $\gamma \geq\gamma_{s}$&
 $\lambda>0$ & Yes& $\lambda<0$ & Yes\\
& & & & & & & & & $0\leq\gamma\leq\gamma_{s}$ &  & $0\leq\gamma\leq\gamma_{s}$ &  \\
\hline $\epsilon=-1$&$\forall\lambda,\forall\gamma$ &
 No & $\forall \lambda, \, \forall \gamma$ & when &
 $\forall\lambda,\forall\gamma$ & when &$\forall\lambda,\,\forall\gamma$ &
 No& $\lambda>0 $ & No & $ \lambda<0 $& No\\
& & & & exists & & exists  & & & $0\leq\gamma\leq\gamma_{s}$ & & $0\leq\gamma\leq\gamma_{s}$ &  \\
\hline
\end{tabular}
\caption{The existence and stability conditions for an expanding
($\epsilon=1$) and a contracting ($\epsilon=-1$) universe,
containing a tachyon field and a fluid with the adiabatic index
$\gamma$. }
\end{table}
\end{widetext}

In addition, we also plot some representative figures (FIG.1 and FIG.2) for the evolution of universe, from which we can see the better picture of the late time acceleration. From FIG.1, we can see that when $\gamma=1>\gamma_{s}$, the evolution of the parameters $x$ and $y$ approach the critical point $({X}_{c}=\lambda Y_{s}/\sqrt{3}\simeq 0.53,$ ${Y}_{c}=Y_{s}\simeq 0.92)$ as $N$ increases. From FIG.2, we can see that in the attractor regime, the equation of state $w_{\phi}\simeq -0.72$ and the density parameter $\Omega_{m}=0$ but $\Omega_{\phi}=1$ which is just tachyon field dominated epoch.

\begin{figure}
\center{
\includegraphics[scale=1]{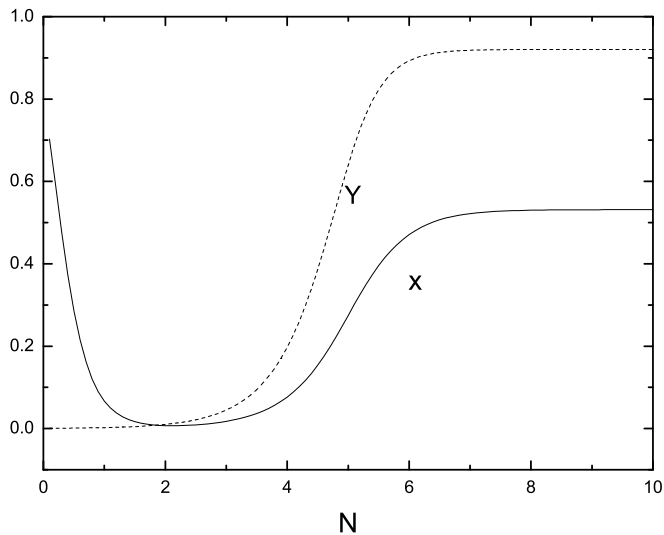}% Here is how to import EPS art
\caption{\label{fig:wide}The evolution of the parameters $x$ and $y$ in the presence of a barotropic fluid with $\gamma=1$. Here we take $\lambda=1$ and the initial conditions $x_{i}=0.8$, $y_{i}=5.0\times 10^{-4}$ when $N=0$.}}
\end{figure}

\begin{figure}
\center{
\includegraphics[scale=1]{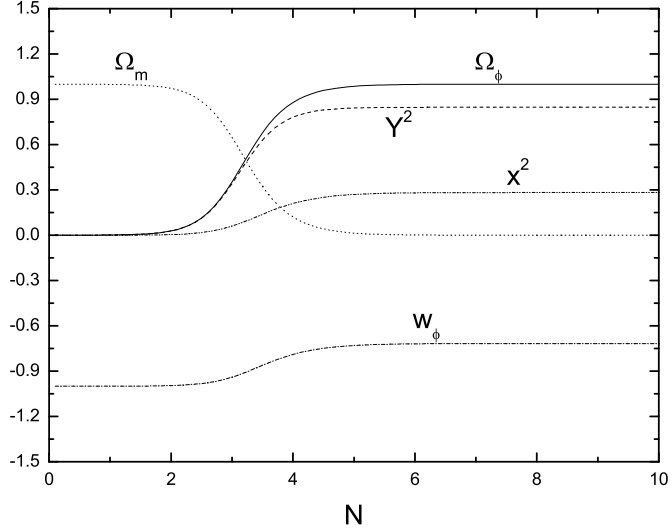}% Here is how to import EPS art
\caption{\label{fig:wide}The evolution of $\Omega_{\phi}$, $\Omega_{m}$, $w_{\phi}$, $x^{2}$ and $y^{2}$ for $\gamma=1$. Here we take $\lambda=1$ and the initial conditions $x_{i}=0.0085$, $y_{i}=0.0085$ when $N=0$.}}
\end{figure}

\section{tachyon field potentials in curved universe}

In the rest of our Letter we focus on the scaling solution given by
Eq.(\ref{ap}), and construct the tachyon field potential which may
give rise to this late time attractor, provided that the specific
forms of $L(\rho)$ is given for spatially curved universe. At
critical points we always have
$Y_{c}=\frac{\sqrt{V}}{\sqrt{\rho}}=constant$, thus $L(\rho)$ can
be described as $L(V)$. Substituting this into  Eq. (\ref{al}), we
can obtain
\begin{equation}\label{az}
\int\frac{d V}{LV^{3/2}}=-\lambda\phi.
\end{equation}

Therefore, given a specific form of $L$, it is possible to derive
the potential for tachyon field in the scaling solution by
integrating Eq.(\ref{az}).

\subsection{Open FRW universe}

In open FRW universe, $L(\rho)$ is given by

\begin{equation}\label{ba}
L(\rho) = \sqrt{1+\frac{3}{\rho a^{2}}},
\end{equation}

At first we assume $\lambda\neq0$ and discuss the form of the
potential in an expanding universe. Then, the special case of
$\lambda=0$ is presented at the end of this subsection.

$\blacksquare$ \textbf{case (i)} \textbf{$\lambda \neq 0 $}

When the universe is expanding ($\epsilon=1$), the solution given
by Eq.(\ref{ap}) is stable for $
\gamma\geq\gamma_{s}=\frac{1}{3}\lambda^{2}Y_{s}^{2}$. By using
the energy conserved equation of tachyon field

\begin{equation}\label{conservedeoftf}
\dot{\rho_{\phi}}+3\frac{\dot{a}}{a}\gamma_{\phi}\rho_{\phi}=0.
\end{equation}

Therefore, we can obtain

\begin{equation}\label{atachyon}
a=a_{(i)}\rho_{\phi}^{\frac{1}{3\gamma_{(i)\phi}}}\rho_{\phi}^{-\frac{1}{3\gamma_{\phi}}}.
\end{equation}

Because this solution is the scalar field dominated scaling solution,
the fluid is absent. Therefore, we can have

\begin{equation}\label{atotal}
a=a_{(i)}\rho^{\frac{1}{3\gamma_{(i)}}}\rho^{-\frac{1}{3\gamma}}.
\end{equation}

With the help of the expression of the effective adiabatic index
$\gamma_{\phi}=X_{c}^{2}=\frac{1}{3}\lambda^{2}Y_{s}^{2}$,
the scale factor $a$ is given by
\begin{equation}\label{bb}
a = A \rho^{-\frac{1}{\lambda^{2}Y_{s}^{2}}}, A =a_{(i)}
\rho_{(i)}^{\frac{1}{\lambda^{2}Y_{s}^{2}}},
\end{equation}
where the subscript ($i$) refers to some initial time.

The correction function $L$ given by Eq.(\ref{ba}) is then
rewritten as

\begin{equation}\label{bc}
L(\rho)=\sqrt{1+\frac{3}{A^{2}}\rho^{\mu}},
\end{equation}
where $ \mu=\frac{2}{\lambda^{2}Y_{s}^{2}}-1$, and for the region of
existence of these solutions, the valid range of $\mu$ is
$-1/3<\mu<\infty$.

At the critical point, $Y_{c}$ is a constant. Substituting
$Y_{s}=\sqrt{V/\rho}$ into Eq. (\ref{az}), the scaling solution
potential will correspond to

\begin{equation}\label{bd}
\int\frac{d
V}{V^{3/2}\sqrt{1+\frac{3}{A^{2}}Y_{s}^{-2\mu}V^{\mu}}}=-\lambda\phi.
\end{equation}

Subsequently, we will classify the asymptotic behavior of the scalar
field potential above in terms of the sign of the parameter $\mu$.
In addition, we also note that the equation (\ref{bd}) is invariant
under the transformation $\lambda \rightarrow -\lambda$ and $\phi
\rightarrow -\phi$. Therefore, our work can be restricted to the
first quadrant without loss of generality.

\begin{enumerate}
\item For $\mu < 0$ $(i.e. \lambda^{2}Y_{s}^{2} >2)$, in an
expanding universe at early times, where the curvature is
negligible, the potential for the tachyon field is obtained as $V
\sim 4\lambda^{-2}\phi^{-2}$. However, at late times, the universe
becomes one dominated by the curvature, the asymptotic form of the
potential is a function like $ V \propto
\lambda^{-\frac{2}{\mu+1}}\phi^{-\frac{2}{\mu+1}}$.

\item For $\mu > 0$ $(i.e. \lambda^{2}Y_{s}^{2} <2)$, we find that
at early times, the potential relating to the curvature dominated
universe can be approximated by $ V \propto
\lambda^{-\frac{2}{\mu+1}}\phi^{-\frac{2}{\mu+1}}$. Once the
curvature is negligible at late times, it can be shown that the
potential is $V \sim 4\lambda^{-2}\phi^{-2}$.

\item For the special case of $\mu = 0$, the correction function
$L$ becomes a constant, and it can be seen as a modified Newton's
gravitational constant in Eq.(\ref{aa}). The evolving background
corresponds to a flat Friedmann universe, and the potential
asymptotically has a form of the late-time attractor $V \sim
\frac{2\sqrt{3}}{3}\phi^{-2}$. Note that in this special case, the
scalar field and the curvature scale together.

\end{enumerate}
We summarize above results in the Case A category of Table II.

However, according to the results of Table I, we know that in a
contracting universe, the scalar field dominated solution is not
stable, but the kinetic dominated solution is stable. We must point
out that in kinetic dominated solution, since $Y_{c}=0$, we have the
potential $V(\phi)=0$.

$\blacksquare$ \textbf{case (ii)} \textbf{$\lambda \approx 0 $}

For the special case of $\lambda=0$, the potential will be a
constant.  The solution described by Eq.(\ref{ap}) is still
applicable and from it we obtain $X_{c} = 0$, $Y_{c} = 1$, which
is nothing but a de-sitter solution, as substituting this into
Eqs.(\ref{ab})and (\ref{ac}) we have $ \dot{\phi}=0 $ and the state
parameter $\omega = -1$. The exact solution can also be obtained
by substituting these into the Eqs.(\ref{aa}), (\ref{ae}) and
(\ref{af}).

From Table I, we can conclude that the tachyon field dominated
solution is stable in the expanding universe, but unstable in a
contracting universe.

\subsection{Closed FRW universe}

Now we turn to obtain the tachyon potential at the stable critical
point for a closed FRW universe. In this case $L(\rho(a))$ is
given by
\begin{equation}\label{be}
L(\rho) = \sqrt{1-\frac{3}{\rho a^{2}}}.
\end{equation}

Obviously, the valid range of the correction function is $1\geq L
\geq 0$. Specially when $L = 0$, the scale factor is a constant,
which neither expanding nor contracting. Therefore, we can assume
at that point, the universe probably experiences a bounce and the
evolution transmits from expanding(contracting) to contracting
(expanding). Let us follow this idea to do the rest of the
analysis. As before, we begin our discussion on the expanding
universe and then analyze the difference between expanding and
contracting universe for the scaling solution.

$\blacktriangledown$ \textbf{case (i)} \textbf{$ \lambda\neq 0 $}

In an expanding universe($\epsilon=1$), the scaling solution given
by Eq.(\ref{ap}) exists for $\gamma \geq \gamma_{s} =
\frac{\lambda^{2}Y_{s}^{2}}{3}$, and the scale factor is still
given by Eq.(\ref{bb}). Then we can rewrite the correction
function as
\begin{equation}\label{bf}
L(\rho)=\sqrt{1-\frac{3}{A^{2}}\rho^{\mu}}.
\end{equation}

Subsequently, using $Y_{c}=\frac{\sqrt{V}}{\sqrt{\rho}}$ and
substituting (\ref{bf}) into Eq.(\ref{az}), we obtain
\begin{equation}\label{bg}
\int\frac{d
V}{V^{3/2}\sqrt{1-\frac{3}{A^{2}}Y_{s}^{-2\mu}V^{\mu}}}=-\lambda\phi.
\end{equation}

As the case in the open FRW universe, we discuss the asymptotic
behavior of the scaling potential by classifying the valid range of
$\mu$ into positive, negative and vanishing regions. We can also
choose to work in the first quadrant without loss of generality.
These are summarized in Case B of Table II.

\begin{enumerate}

\item For $\mu>0$  (i.e. $\lambda^{2}Y_{s}^{2}<2$), at early time
of an expanding universe, we have
$a^{-\lambda^{2}Y_{s}^{2}}<a^{-2}$. However, the universe cannot be
dominated by the curvature as the correction function is bounded in
a closed universe, otherwise the right-hand side of the Friedmann
equation (\ref{aa}) would be negative. Consider a pragmatic point,
the curvature is just subdominant to the scalar field. When
$L(\rho)\rightarrow 0 $, the total energy density has a maximum
$\rho\rightarrow\rho_{max}\equiv(\frac{A^{2}}{3})^{1/\mu}$, and the
universe has a constant scale factor. From Eq.(\ref{ah}), we
conclude that the potential is almost a constant with $V\approx
Y_{c}^{2}(\frac{A^{2}}{3})^{1/\mu}$. Subsequently, the universe
starts to expand, the tachyon field gradually become dominated over
the curvature. At last, one can neglect the contribution of the
curvature. Therefore, the universe can be considered a flat FRW
spacetime, and the asymptotic form of the potential become
$V\sim4\lambda^{-2}\phi^{-2}$.

\item For $ \mu<0 $ (i.e. $\lambda^{2}Y_{s}^{2}>2$), the universe
dominated by the tachyon field starts to expand from an
approximate flat FRW spacetime, and the scaling solution potential
has the form $V\sim4\lambda^{-2}\phi^{-2}$. As time goes on, the
total energy density will reach the maximal value, $
\rho\rightarrow\rho_{max}\equiv(\frac{A^{2}}{3})^{1/\mu}$, where
the contribution of the curvature and the scalar field are almost
equal. The scale factor does not change as $H\rightarrow 0$, and
the potential is almost a constant $V\approx
Y_{c}^{2}(\frac{A^{2}}{3})^{1/\mu}$. After this turning point, the
universe changes from expanding to collapsing. From Table I, we
see that in a contracting universe, the late time scalar field
scaling solutions are unconditionally unstable. Once the universe
starts to collapse, the solution will asymptote towards the
kinetic dominated solution with $\gamma_{\phi}=1$.

\item For $\mu=0 $, the situation is similar to the open universe
scenario. The correction function $L$ becomes a constant. The
potential has the form $V \sim \frac{2\sqrt{3}}{3}\phi^{-2}$,
where the contributions of the scalar field and curvature to the
Friedmann equation scale together.

\end{enumerate}

As mentioned in the subsection for open universe, for a contracting
universe if the kinetic dominated solution exists they will be
unconditionally stable, and the potential $V(\phi)=0$.

$\blacktriangledown$ \textbf{case  (ii)} \textbf{$\lambda \approx 0
$}

Similar to our description of an open universe scenario, this case
corresponds to a constant potential, and only the fixed point
described by (\ref{ap}) exists. For $\lambda=0$, which results in
an expanding universe, the attractor solution reduces to $X_{c} =
0$ and $Y_{c} = 1$, describing an exact de-Sitter solution. We
notice that the assumption of taking $\phi$ as a monotonically
varying function of time breaks down if $X_{c}\propto \dot{\phi}=
0$.

Furthermore, based on our previous analysis it is worth noting
that when $\lambda\approx 0$, the critical point (\ref{ap}) is
unstable in a contracting universe, but becomes a stable one in an
open universe.

Finally, from the above analysis, we note that the evolution of
the universe including the curvature term is same as the case
without the curvature term \cite{tachyondeEJC}. However, due to
the curvature term, the required potential is different. As shown in \cite{tachyondeEJC},
in the case without curvature term, the required potential is always $V(\phi)\sim \phi^{-2}$.
But when the curvature term is taken into account,  the potential has various form in the different epochs of
the universe.

\begin{widetext}
\begin{center}
\begin{table} [h!]
\begin{tabular} { |c|c|c|c|c|c|c| }
\hline
 & \multicolumn{3}{|c|}{\em Case A} & \multicolumn{3}{|c|}{\em Case B} \\
\hline
 & $\mu<0$ & $\mu=0$ & $\mu>0$ & $\mu<0$ & $\mu=0$ & $\mu>0$ \\
\hline \hline
 & $V \sim \phi^{-2}$ & $V \sim \phi^{-2}$ & $V \sim \phi^{-\frac{2}{\mu+1}}$ & $V \sim \phi^{-2}$ & $V \sim \phi^{-2}$ & $V \sim Y_{c}^{2}(\frac{A^{2}}{3})^{1/\mu}$   \\
\raisebox{1.5ex} {Early times} & negligible & scaling & curvature & negligible & scaling & curvature \\
 & curvature & curvature & dominated & curvature & curvature & only just \\
 & & & & & & subdominant \\
 \hline
 & $V \sim \phi^{-\frac{2}{\mu+1}}$ & $V \sim \phi^{-2}$ & $V \sim \phi^{-2}$ & $V \sim  Y_{c}^{2}(\frac{A^{2}}{3})^{1/\mu}$ & $V \sim \phi^{-2}$ & $V \sim \phi^{-2}$  \\
\raisebox{1.5ex} {Late times} & curvature & scaling & negligible & curvature & scaling & negligible \\
 & dominated & curvature & curvature & only just & curvature & curvature \\
 & & & & subdominant & & \\
\hline
\end{tabular}
\caption{This table summarizes the asymptotic behavior of an
expanding ($\epsilon = 1$) universe described by the scaling
solutions. {\em Case A} refers to open FRW universe, and {\em Case
B} corresponds to closed FRW universe.} \label{Open_FRW}
\end{table}
\end{center}
\end{widetext}

\section{Conclusion and discussion}

In this Letter we have investigated the dynamics of a tachyon field
in FRW universe with spatial curvature. Following the scheme
presented in \cite{SS312,CSQ}, we denoted the modification of the
cosmological equation due to the spatial curvature by a general
function $L(\rho)$, and then derived the conditions under which the
system can enter a scaling solution. In particular, we obtain an
attractor solution, where the tachyon field dominates over the fluid
and the kinetic energy of the field scales with its potential
energy. Furthermore, given the modification function for an open and
closed universe respectively, we discussed the form of the scalar
potential which gives rise to the late time attractor solution. For
spatially open universe, we conclude that in regions where the
curvature is negligible, the asymptotic form of the potential is
$V\propto\phi^{-2}$, while in regions where it is dominant, the
approximated potential will be $V \propto\phi^{-\frac{2}{\mu+1}}$.
As far as the spatially closed universe is concerned, we find that
the curvature is not allowed to be dominant among the total energy
density of the universe, but only subdominant to the scalar field
density, where the potential is a constant as $V\approx
Y_{c}^{2}(\frac{A^{2}}{3})^{1/\mu}$. When the curvature becomes
negligible, the potential can be approximated  by $ V \sim \phi^{-2}
$. Moreover, it was noticed that in a contracting closed universe,
the scalar dominated solutions are not stable, and the kinetic
dominated solution will be a late time attractor if it exists.

Comparing with the quintessence scalar field dark energy model with
spatial curvature \cite{CSQ}, the fluid-scalar field scaling solutions
is absent in the tachyon dark energy model. In addition, in the
quintessence scalar field dark energy model, there are the two forms
of the potential: the exponential potential and the power-law potential.
But in the tachyon dark energy model, only the power-law potential is required.

In this Letter we only consider the case that the scalar field and
the fluid do not interact with each other. It is a very
interesting question if such an interaction term can be introduced
in order to obtain a fluid-tachyon field scaling solutions in
curved FRW universe. Effort has been made in present framework,
unfortunately, the answer seems to be negative. However, it should
be very desirable if we consider adding a Gauss-Bonnet coupling as
in \cite{SS7}. We expect to make further progress along this
direction.

\section*{Acknowledgement}

We are grateful to Prof. Yi Ling and Da-Zhu Ma for
helpful discussions. This work is partly supported by NSFC(Nos.10663001,10875057),
JiangXi SF(Nos. 0612036, 0612038), Fok Ying Tung Eduaction
Foundation(No. 111008) and the key project of Chinese Ministry of
Education(No.208072). We also acknowledge the support by the
Program for Innovative Research Team of Nanchang University.

\end{document}